\documentclass[twocolumn,aps,pra,showpacs]{revtex4}

\usepackage{amsmath}
\usepackage{amssymb}
\usepackage{amsfonts}
\usepackage{graphicx}

\begin{document}

\title{Reply to "Comment on `Stimulated Raman adiabatic passage from an atomic to a molecular Bose-Einstein condensate'"}

\author{P. D. Drummond}

\author{K. V. Kheruntsyan}

\affiliation{ARC Centre of Excellence for Quantum-Atom Optics, Department of Physics, University of Queensland, Brisbane, Qld 4072, Australia}

\author{D. J. Heinzen}

\author{R. H. Wynar}

\affiliation{Department of Physics, University of Texas, Austin, Texas 78712}

\date{\today}


\begin{abstract}
In the Comment by M. Mackie \textit{et al.} [arXiv: physics/0212111 v.4], the authors suggest that the molecular conversion efficiency in atom-molecule STIRAP can be improved by lowering the initial atomic density, which in turn requires longer pulse durations to maintain adiabaticity. Apart from the fact that the mean-field approximation becomes questionable at low densities, we point out that a low-density strategy with longer pulses has several problems. It generally requires higher pulse energies, and increases radiative losses. We also show that even within the approximations used in the Comment, their example leads to no efficiency improvement compared to our high-density case. In a more careful analysis including radiative losses neglected in the Comment, the proposed  strategy gives almost no conversion owing to the longer pulse durations required.
\end{abstract}
\maketitle
The authors of Ref. \cite{MM-JJ-Comment} revisit our earlier work on stimulated Raman adiabatic passage (STIRAP) from an
atomic to a molecular Bose-Einstein condensate \cite{Drummond-etal-STIRAP}. They suggest that the disruptive role of dephasing
due to particle-particle interactions can be reduced by lowering the initial atomic condensate density. This strategy requires
a simultaneous increase of the duration of the Raman pulses. As a result the authors claim to obtain improved conversion
efficiency, which reaches about 51\% for the optimized density. This is about 5\% larger than the highest efficiency of $\sim$46\%
found in Ref. \cite{Drummond-etal-STIRAP}, and is achieved by using a 100 times lower atomic density and about 100 times
longer pulse duration.

Apart from the fact that mean-field theory becomes questionable for describing low density condensates, we start by emphasizing
that any comparison of efficiency is only meaningful within a well-defined situation of equal atom numbers and laser powers (or pulse energies) \cite{Laser-footnote}. This is because the optimum efficiency is a strong function of maximum Rabi frequency. Given these constraints, together with increased radiative losses at longer pulse durations,
we show here that there are three significant problems with the results of Ref. \cite{MM-JJ-Comment}.

\emph{Laser power.} The first problem is that a low density strategy requires a higher laser power to obtain an equal laser intensity, since
the laser waist size must be enlarged. This is due not only to the larger size of the condensate, for fixed atom number,
but also to the fact that the experiment becomes more sensitive to the inhomogeneity in the AC Stark shifts from the photoassociation
laser beams, which set a practical lower limit on the beam waists. If such high power lasers to implement the low density
strategy of Ref. \cite{MM-JJ-Comment} are available, they can simply be utilized to increase the maximum Rabi frequency
in the original proposal of Ref. \cite{Drummond-etal-STIRAP} -- which greatly increases the conversion efficiency. Using
fewer atoms, as suggested in the Comment, invalidates the comparison.

\textit{Parameter normalization and optimization.} The next problem is that equations (1a)-(1c) of Ref. \cite{MM-JJ-Comment} and the
parameters used do not correspond to those of Ref. \cite{Drummond-etal-STIRAP}. The difference is in the starting value of the parameter
$\chi_{0}$ and the fact that the respective equations of Ref. \cite{Drummond-etal-STIRAP} contain an extra factor $1/\sqrt{2}$
in front of this coupling term. Hence, the comparison is made
between different Rabi frequencies, not just a lower atomic density.

Thus, the results of the Comment should be compared with uniform trap results of Ref. \cite{Drummond-etal-STIRAP} obtained
with $\sqrt{2}$ times larger Rabi frequency $\Omega_{1}$ (see Eqs. (6) of Ref. \cite{Drummond-etal-STIRAP}). In this case,
simulation of the respective equations of Ref. \cite{Drummond-etal-STIRAP} with a factor of $\sqrt{2}$ larger peak value of
$\Omega_{1}$, zero two-photon detuning, and the same original values for the remaining parameters (i.e. higher atomic density $\rho_{0}=4.3\times10^{14}$
cm$^{-3}$, a pulse duration of $T=0.2$ ms, and a 50 times larger peak value of $\Omega_{2}$ than $\Omega_{1}$) gives $67$\% conversion efficiency. This is higher than the maximum efficiency of $51\%$ found in the Comment \cite{MM-JJ-Comment}.

In other words, the claimed improvement in conversion is obtained not just by altering the initial atomic density
and pulse duration, but by changing other parameter values as well. When directly comparable Rabi frequencies are used, the
relative efficiency is in fact worse in the suggested low-density strategy.

We also point out that the conversion efficiency of $16$\% seen in
Fig. 1 of the Comment at $\rho_{0}=4.3\times10^{14}$ cm$^{-3}$
should not be confused with our result of $67$\% at the same
density. The reason for this seeming disagreement is that the
authors of the Comment use essentially a one-parameter (density)
optimization. This means that other parameter values like pulse
duration are not necessarily optimized at all densities. For
example, the pulse duration is set to $T=5\times10^{3}/\chi_{0}$
in all cases, which for $\rho_{0}=4.3\times10^{14}$ cm$^{-3}$
gives $T=2.4$ ms. In contrast to this, our original optimization
procedure is carried out with respect to three parameters -- the
pulse duration, pulse offset, and the two-photon detuning. The
$67$\% efficiency obtained in our case uses an optimum pulse
duration of $T=0.2$ ms, which is much shorter than used in the
Comment at this density \cite{Optimization}.

It should be noted that $67$\% efficiency is higher than the maximum of $46$\% we found previously \cite{Drummond-etal-STIRAP}. This is simply due to the improved adiabaticity following obtained at higher Rabi frequency.

\textit{Radiative losses}. We now address the most detrimental outcome of the longer pulse durations employed in the Comment. As the timescale for STIRAP is increased, incoherent radiative losses become increasingly important due to additional radiative couplings that are ignored in the model of Ref. \cite{MM-JJ-Comment}.
We show here that once included in the model, they can lead to dramatic losses of the ground-state molecules. Our results
for typical parameter values show almost zero final population. Thus, in the suggested case of $^{87}$Rb, the proposed
strategy results in a dramatic reduction rather than improvement in the conversion efficiency.

There are two types of additional radiative couplings that are neglected in the Comment: (1) couplings that occur even within the simple three-level model, and (2)
couplings to other vibrational levels in the excited molecular potential.

The couplings within the three-level model included in our original paper \cite{Drummond-etal-STIRAP,Comment1}
are due to non-resonant interactions of the two Raman lasers. These
lead to the following, more complete equations:
\begin{align}
i\dot{a} & =\left(\frac{1}{2}\Delta+\Lambda_{aa}|a|^{2}+\Lambda_{ag}|g|^{2}\right)a-\chi a^{\ast}b\nonumber \\
 & -\tilde{\chi}e^{-i\omega_{12}t}a^{\ast}b,\label{a}\end{align}
\begin{align}
i\dot{b} & =\left(\delta-\frac{i}{2}\gamma_{s}\right)b-\frac{1}{2}\left(\chi a^{2}+\Omega g\right)\nonumber \\
 & -\frac{1}{2}\left(\tilde{\chi}e^{i\omega_{12}t}a^{2}+\tilde{\Omega}e^{-i\omega_{12}t}g\right),\label{b}\end{align}
\begin{align}
i\dot{g} & =\left(\Lambda_{ag}|a|^{2}+\Lambda_{gg}|g|^{2}\right)g- \frac{1}{2}\Omega b\nonumber \\
 & -\frac{1}{2}\tilde{\Omega}e^{i\omega_{12}t}g.\label{g}\end{align}
 Here, \begin{align*}
\chi(t) & =\chi_{0}\exp\left[-(t-D_{1})^{2}/T^{2}\right],\\
\tilde{\chi}(t) & =\tilde{\chi}_{0}\exp\left[-(t-D_{2})^{2}/T^{2}\right],\\
\Omega(t) & =\Omega_{0}\exp\left[-(t-D_{2})^{2}/T^{2}\right],\\
\tilde{\Omega}(t) & =\tilde{\Omega}_{0}\exp\left[-(t-D_{1})^{2}/T^{2}\right],\end{align*}
 while $\omega_{12}=\omega_{1}-\omega_{2}$ is the frequency difference between the two Raman lasers, $\Delta$ ($\delta$)
is the two-photon (intermediate) detuning, and $\Lambda_{ij}$ represent atom-atom, atom-molecule, and molecule-molecule
$s$-wave scattering interactions. The complex amplitudes $a$, $b$, and $g$ represent the atoms, excited molecules, and
stable molecules in the ground potential, respectively.

For the benefits of the readers, we use the same notations as in Ref. \cite{MM-JJ-Comment}. Here, the density-dependent
coupling $\chi$ describes free-bound transitions between atom pairs and excited molecules due to the Raman laser at frequency
$\omega_{1}$, which is the second pulse in the counterintuitive STIRAP sequence. The Rabi frequency $\Omega$ describes
the bound-bound transitions between excited and ground-state molecules due the first Raman pulse at frequency $\omega_{2}$.
In addition to these couplings, which are the primary transitions in any STIRAP scheme, the above equations include the coupling
of atoms to excited molecules due to the $\omega_{2}$-laser, and similarly the coupling of excited and ground-state molecules
due to the $\omega_{1}$-laser. The respective coupling constants are $\tilde{\chi}$ and $\tilde{\Omega}$, and in the case
of $\Delta=\delta=0$ these non-primary or `partner' transitions are detuned by $\omega_{12}$.

Explicit parameter values here are taken as in Ref. \cite{MM-JJ-Comment} to make the comparison valid: $\Delta=0$, $\delta=\chi_{0}$, $\gamma_{s}=7.4\times10^{4}$
s$^{-1}$, $\Lambda_{aa}=213$ s$^{-1}$, $\Lambda_{gg}=107$ s$^{-1}$, $\Lambda_{ag}=-277$ s$^{-1}$, and the initial
atomic density is $\rho=4.3\times10^{12}$ cm$^{-3}$ so that $\chi_{0}=2.1\times10^{5}$ s$^{-1}$. In addition, $\Omega_{0}=50\chi_{0}=10^{7}$
s$^{-1}$, $T=5\times10^{3}/\chi_{0}=24$ ms, $D_{1}=4.5T$, and $D_{2}=2.5T$. Finally, we take $\tilde{\chi}_{0}=10^{4}$
s$^{-1}$, $\tilde{\Omega}_{0}=2.3\times10^{8}$ s$^{-1}$, and $\omega_{12}=-5.1\times10^{10}$ s$^{-1}$ which are close
to the typical calculated values corresponding to the spectroscopically most favorable case treated in Ref. \cite{Drummond-etal-STIRAP}.
The values of the
new parameters $\tilde{\chi}_{0}$ and $\tilde{\Omega}_{0}$ are chosen in a favorable manner \cite{Comment2}, and our conclusions
would remain valid if the above mentioned `missing' factors of $1/\sqrt{2}$ were restored self-consistently.

Using the above parameters and simulating Eqs.\ (\ref{a})-(\ref{g}), gives a final population of the ground-state molecules
of $|g|^{2}\simeq0$, while the peak value during the pulse sequence reaches only $|g|^{2}\simeq2.5\times10^{-6}$. This
implies essentially zero conversion efficiency.

The reason for this dramatic result is that the newly formed ground-state molecules are being still illuminated by the laser
$\omega_{1}$ during the second Raman pulse. As a result, they experience radiative losses at a rate of \begin{equation}
\Gamma_{eff}\equiv\frac{\gamma_{s}}{4}\left|\frac{\tilde{\Omega}_{0}}{\omega_{12}}\right|^{2}=376\;\text{s}^{-1}.\label{Gamma-eff}\end{equation}
 Thus, the characteristic time scale for losses is $1/\Gamma_{eff}\sim2.6$ ms, which is much smaller than the pulse duration
$T\simeq24$ ms employed in Ref. \cite{MM-JJ-Comment}. These radiative losses are negligible for the much shorter (sub-millisecond)
pulses treated in Ref. \cite{Drummond-etal-STIRAP}. In this case, $\Gamma_{eff}T\ll1$ and the role of the non-primary transitions
is negligible.

One might argue that targeting lower-lying vibrational levels in the ground molecular potential would give larger detuning
$\omega_{12}$, thus making $\Gamma_{eff}$ smaller. This approach, however, suffers from the fact the respective bound-bound
Franck-Condon overlap integrals typically become smaller and hence give even smaller values of the Rabi frequency $\Omega_{0}$.
In addition, a detailed multilevel analysis reveals that these lower lying levels do not correspond to the most favorable
case, once we take into account the entire set of necessary conditions for efficient conversion (see Eqs. (48)-(54) and the
typical parameter values in Table IV of Ref. \cite{Drummond-etal-STIRAP}). The reason for this is that increasing the detuning
$\omega_{12}$ will eventually bring the frequency of one of the lasers (or both) to a nearby resonance in the excited potential,
thus giving rise to additional induced losses just as in Eq. (\ref{Gamma-eff}).

The induced molecular losses due to the couplings to other vibrational levels in the excited potential can be modeled by
the following additional term in the right-hand side of Eq. (\ref{g}): \begin{equation}
i\dot{g}=(\;...)-\frac{i}{2}\Gamma_{2}g,\label{g-loss}\end{equation}
 where\begin{equation}
\Gamma_{2}=\Gamma_{2}^{(1)}e^{-2(t-D_{1})^{2}/T^{2}}+\Gamma_{2}^{(2)}e^{-2(t-D_{2})^{2}/T^{2}},\end{equation}
 and the expressions for $\Gamma_{2}^{(i)}$ are given by Eq.\ (43) of Ref. \cite{Drummond-etal-STIRAP}. Here, the most
disruptive loss coefficient is $\Gamma_{2}^{(1)}$ which is due to the bound-bound transitions during the second Raman pulse
(laser $\omega_{1}$). We take $\Gamma_{2}^{(1)}\simeq400$ s$^{-1}$ here, which is more favorable than the value obtained
in Ref. \cite{Drummond-etal-STIRAP}. This gives a characteristic time scale for losses $1/\Gamma_{2}^{(1)}\simeq2.5$ ms,
which is much shorter than the pulse duration $T\simeq24$ ms. Not surprisingly, the simulation of Eqs.\ (\ref{a})-(\ref{g})
with the additional loss term, Eq.\ (\ref{g-loss}), and with $\tilde{\chi}_{0}$ and $\tilde{\Omega}_{0}$ set to zero,
gives again almost zero final population of the ground-state molecules.

An examination of the conditions stated in Eqs. (48)-(54) of Ref. \cite{Drummond-etal-STIRAP} and of the characteristic
values of the respective coefficients in Table IV reveals that these results could have been expected. The entries in lines
5 and 12 of Table IV give the typical values of $\Gamma_{2}^{(1)}$ and $\Gamma_{eff}$. With pulse durations of $T\simeq24$
ms, these coefficients do not satisfy the conditions $\Gamma_{2}^{(1)}T\ll1$ and $\Gamma_{eff}T\ll1$, and therefore the
induced losses are not negligible. Similarly, even the induced atomic loss coefficient $\alpha^{(1)}=51$ s$^{-1}$ gives
$\alpha^{(1)}T\simeq1.2$ (with $T=24$ s$^{-1}$) and hence it cannot be neglected either.

We should note that all of the calculations we presented have been carried out within a simplified model of the molecular
structure, which assumes an $^{87}$Rb$_{2}$ molecule with a single ground and single excited electronic state. We include
vibrational, but neglect rotational and hyperfine structure. As such, our conclusions -- while correct qualitatively -- can be further improved for quantitative purposes. Nevertheless, in contrast to the calculations of Ref. \cite{MM-JJ-Comment}, ours do include the important physical
effects in a reasonably realistic way, within the limitations of mean-field theory. We feel that the general conclusions
we have presented, such as the poor scaling of the molecular conversion efficiency with decreasing density, and the lack
of very substantial improvement in conversion with increasing molecular binding energy, are likely to be present
in any STIRAP experiment with alkali dimers.

To summarize, the photoassociation strategy employed in Ref. \cite{MM-JJ-Comment} requires higher laser powers and
pulse energies. Even neglecting losses, it does not lead to improvements in efficiency relative to the original proposal,
provided the maximum Rabi frequencies are kept equal. It also fails to capture the physics of induced radiative losses relevant
for long pulse durations. The proposal of using lower atomic densities and longer pulses for achieving higher conversion
efficiencies in atom-molecule STIRAP gives in fact the opposite effect, once these additional loss channels are taken into
account. We find that with typical parameter values, a more complete model results in almost zero conversion, under the proposed
conditions.

P.D. and K.K. acknowledge the ARC for the support of this work. R. W. and D. H. acknowledge the support of the NSF, the R. A. Welch Foundation, and the ONR Quantum Optics Initiative.


\begin{thebibliography}{1}
\bibitem{Laser-footnote}The Rabi frequencies chosen in the original proposal would correspond approximately to a $100~\mu$m waist and $1$ W power
per Raman channel, typical of current cw Ti:sapphire laser beams. If the beam waist size is much smaller than this, then
its AC Stark shift becomes very inhomogeneous, which creates new problems.
\bibitem{Optimization}It is possible that a better optimization procedure might work better at lower densities, within the model employed in the Comment. However, we have not investigated this in detail, as the low density strategy will have to necessarily employ longer pulse durations to maintain adiabaticity, in which case the model itself becomes physically unrealistic due to the neglected induced radiative losses.
\bibitem{Comment1}Eq. (29) of Ref. \cite{Drummond-etal-STIRAP} contains a typographical error. Here, the exponential term $\exp(-i\omega_{12}t)$
in the first line should read as $\exp(i\omega_{12}t)$. The same term in Eq. (57) is correct. In addition, in
the caption to Table III, $|I_{1,3}|=0.1$ should be replaced by $|I_{2,3}|=0.1$, and the quoted values of $\delta$ should refer to $-\delta$ instead. The value of $|I_{2,3}|$ is correct elsewhere
in the text, and these typographical errors do not affect the results presented in Ref. \cite{Drummond-etal-STIRAP}.
\bibitem{Comment2}On evaluating the values of the parameters $\tilde{\chi}_{0}$ and $\tilde{\Omega}_{0}$, one has to pay attention to the
fact that these have to be calculated in conjunction with $\chi_{0}$ and $\Omega_{0}$ and that they can not be varied independently.
The reason is that these couplings originate from the values of the bare electronic Rabi frequencies for the two transitions,
$\Omega_{1}^{(el,0)}$ and $\Omega_{2}^{(el,0)}$, and the free-bound and bound-bound Franck-Condon overlap integrals, $I_{1,3}$
and $I_{2,3}$, using the notations of Ref. \cite{Drummond-etal-STIRAP}. Accordingly, $\chi_{0}\propto\Omega_{1}^{(el,0)}I_{1,3}$,
$\Omega_{0}\propto\Omega_{2}^{(el,0)}I_{2,3}$, while $\tilde{\chi}_{0}\propto\Omega_{2}^{(el,0)}I_{1,3}$ and $\tilde{\Omega}_{0}\propto\Omega_{1}^{(el,0)}I_{2,3}$.
\end{thebibliography}
\end{document}